\pdfoutput=1

\documentclass[11pt,twoside,a4paper,cmspaper,final,collab]{cms-tdr}
\def\svnVersion{493569P}\def\cmsCernNoTag{CERN-EP-2018-280}\def\cmsCernDate{\today}\def\cmsMessage{Published in the Journal of High Energy Physics as \href{http://dx.doi.org/10.1007/JHEP04(2019)015}{\doi{10.1007/JHEP04(2019)015.}}}

\begin{document}\cmsNoteHeader{EXO-18-004}

\hyphenation{had-ron-i-za-tion}
\hyphenation{cal-or-i-me-ter}
\hyphenation{de-vices}
\RCS$HeadURL: svn+ssh://svn.cern.ch/reps/tdr2/papers/EXO-18-004/trunk/EXO-18-004.tex $
\RCS$Id: EXO-18-004.tex 493559 2019-04-08 10:51:33Z knam $

\newlength\cmsFigWidth
\ifthenelse{\boolean{cms@external}}{\setlength\cmsFigWidth{0.85\columnwidth}}{\setlength\cmsFigWidth{0.4\textwidth}}
\ifthenelse{\boolean{cms@external}}{\providecommand{\cmsLeft}{upper\xspace}}{\providecommand{\cmsLeft}{left\xspace}}
\ifthenelse{\boolean{cms@external}}{\providecommand{\cmsRight}{lower\xspace}}{\providecommand{\cmsRight}{right\xspace}}

\newcommand\lstar{\ensuremath{\ell^*}\xspace}
\newcommand\estar{\ensuremath{\Pe^*}\xspace}
\newcommand\mustar{\ensuremath{\Pgm^*}\xspace}
\newcommand\lstartollg{\ensuremath{\lstar\to\ell\Pgg}\xspace}
\newcommand\meqL{\ensuremath{\Lambda=\mlstar}\xspace}

\newcommand\mll{\ensuremath{{m}_{\ell\ell}}\xspace}
\newcommand\mlstar{\ensuremath{m_{\ell^*}}\xspace}
\newcommand\melstar{\ensuremath{m_{\Pe^*}}\xspace}
\newcommand\mmulstar{\ensuremath{m_{\Pgm^*}}\xspace}
\newcommand\llg{\ensuremath{\ell\ell\Pgg}\xspace}
\newcommand\eeg{\ensuremath{\Pe\Pe\Pgg}\xspace}
\newcommand\mmg{\ensuremath{\Pgm\Pgm\Pgg}\xspace}
\newcommand\Mmin{\ensuremath{m_{\ell\Pgg}^\text{min}}\xspace}
\newcommand\Mmax{\ensuremath{m_{\ell\Pgg}^\text{max}}\xspace}

\newcommand\Zgamma{DY+$\Pgg$\xspace}
\newcommand\ttgamma{$\ttbar$+$\Pgg$\xspace}
\newcommand\WW{\ensuremath{\PW\PW}\xspace}
\newcommand\WZ{\ensuremath{\PW\PZ}\xspace}
\newcommand\ZZ{\ensuremath{\PZ\PZ}\xspace}
\newcommand\WWgamma{\ensuremath{\PW\PW\Pgg}\xspace}
\newcommand\WZgamma{\ensuremath{\PW\PZ\Pgg}\xspace}
\newcommand\ZZgamma{\ensuremath{\PZ\PZ\Pgg}\xspace}

\newcommand\Ndata{\ensuremath{N_\text{data}}\xspace}
\newcommand\Nprompt{\ensuremath{N_\text{prompt}}\xspace}
\newcommand\Njet{\ensuremath{N_\text{jet}}\xspace}
\newcommand\Ae{\ensuremath{A\,\epsilon_\text{sig}}\xspace}

\providecommand{\cmsTable}[1]{\resizebox{\textwidth}{!}{#1}}
\providecommand{\NA}{\ensuremath{\text{---}}}

\cmsNoteHeader{EXO-18-004}
\title{Search for excited leptons in \llg final states in proton-proton collisions at $\sqrt{s}=13\TeV$}

\date{\today}

\abstract{
A search is presented for excited electrons and muons in \llg final states at the LHC.
The search is based on a data sample corresponding to an integrated luminosity of 35.9\fbinv
of proton-proton collisions at a center-of-mass energy of 13\TeV, collected with the CMS detector in 2016.
This is the first search for excited leptons at $\sqrt{s}=13\TeV$.
The observation is consistent with the standard model background prediction,
and the most stringent exclusion limits to date are set on the excited lepton mass and the compositeness scale, at 95\% confidence level.
Excited electrons and muons are excluded for masses below 3.9 and 3.8\TeV, respectively, under the assumption
that the excited lepton mass equals the compositeness scale.
The best observed limit on the compositeness scale is obtained with an excited lepton mass of around 1.0\TeV,
excluding values below 25\TeV for both excited electrons and muons.
}

\hypersetup{
pdfauthor={CMS Collaboration},
pdftitle={Search for excited leptons in llgamma final states in proton-proton collisions at sqrt(s)=13 TeV},
pdfsubject={CMS},
pdfkeywords={CMS, physics, excited leptons}}

\maketitle

\section{Introduction}

The standard model (SM) provides a very precise description of various phenomena in particle physics observed over the last half century.
Notwithstanding its huge success, it does not explain the origin of the mass hierarchy and the number of generations of quarks and leptons.
As an attempt to answer such fundamental questions, compositeness of quarks and leptons is introduced in
many models~\cite{Greenberg:1974qb, Pati:1975md, Greenberg:1980ri, compositeness1, Eichten1982, Eichten:1983hw, harari, Baur90, qstar1, qstar2}.
These compositeness models suggest that quarks and leptons are themselves made of fundamental constituents
that are bound by a new strong interaction with a characteristic energy scale $\Lambda$ (called the compositeness scale).

An important prediction of compositeness models is the existence of excited states of quarks and leptons.
In proton-proton ($\Pp\Pp$) collisions, excited fermions could be produced via contact interactions (CI) and
decay either through SM gauge interactions or via CI to SM fermions.

\begin{figure}[h!]
\centering
\includegraphics[width=0.45\textwidth]{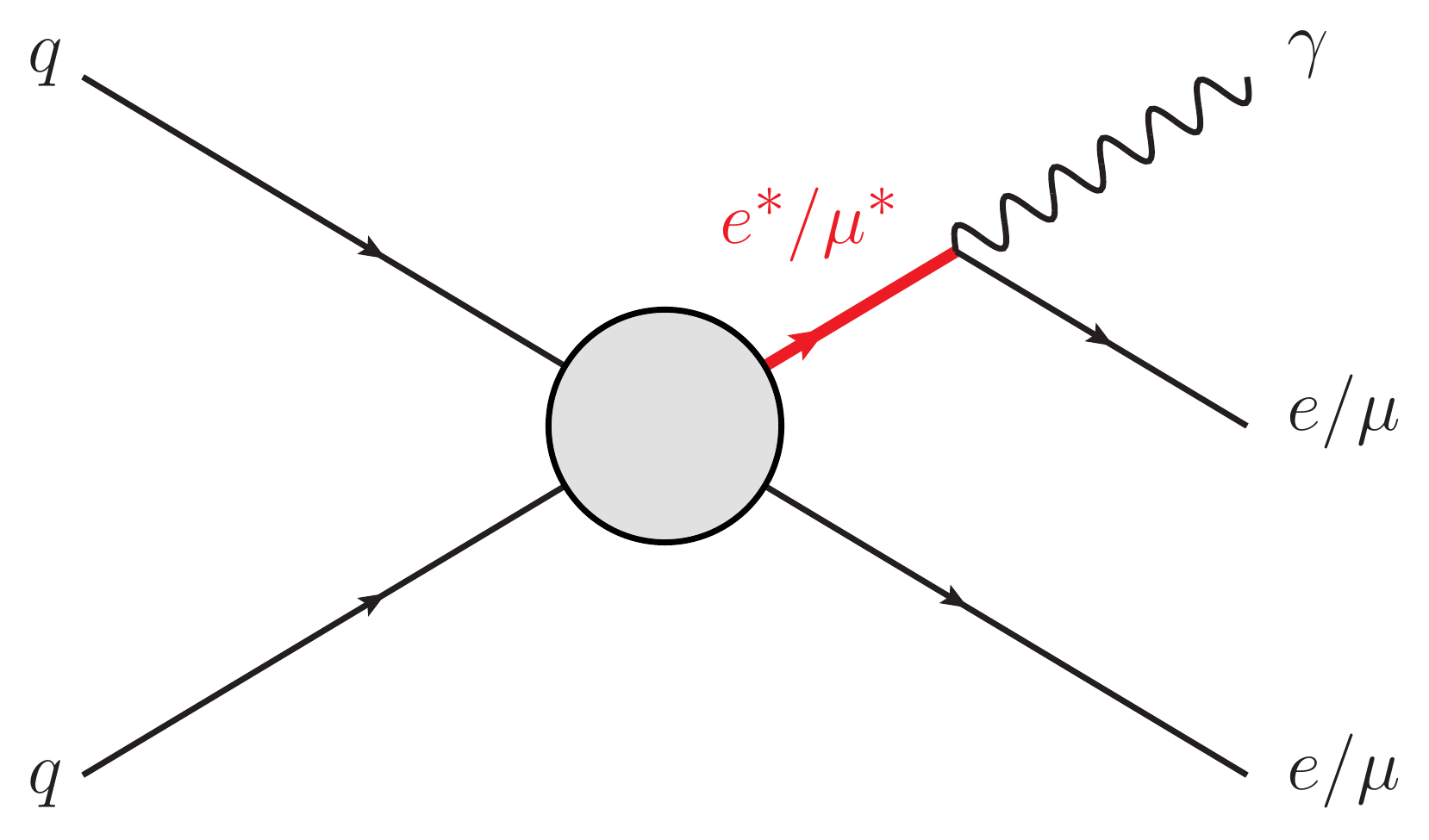}
\caption{\label{fig:Feynman} The Feynman diagram of the production of excited leptons in \llg final states.}
\end{figure}

This paper presents a search for excited leptons ($\lstar=\estar,$ \mustar) in \llg ($\ell=\Pe,$ \Pgm)
final states where the excited lepton decays to a SM lepton and a photon (\lstartollg) as illustrated in Fig.~\ref{fig:Feynman}.
A clear signature of a same-flavor (SF) lepton pair and a photon allows highly efficient signal selection.
However, there is an ambiguity in reconstructing the excited lepton mass because of the two possible pairings of a lepton and the photon.
For this search, information of both invariant masses is used to discriminate the excited lepton signal from SM background processes.

We consider a benchmark model based on the formalism described in Ref.~\cite{Baur90}.
The effective Lagrangian of four-fermion CI, the main production mechanisms of excited leptons at the LHC, is given by
\begin{equation}
\mathcal{L}_{\mathrm{CI}} = \frac{g_{*}^{2}}{2\Lambda^{2}}j^{\mu}j_{\mu},
\end{equation}
where $g_{*}$ is the coupling constant, which is set to be equal to $\sqrt{4\pi}$ in the model,
and $j_{\mu}$ represents the fermion currents for SM fermions and their excited states, neglecting right-handed currents.
In addition, excited lepton decays via SM gauge interactions are described by the corresponding Lagrangian
\begin{equation}
\mathcal{L}_{\mathrm{gauge}} = \frac{1}{2\Lambda}\overline{\ell}_{R}^{*}\sigma^{\mu\nu} \left(gf \frac{\tau}{2} {W}_{\mu\nu} + g^{\prime}f^{\prime}\frac{Y}{2}B_{\mu\nu}
\right) \ell_{L} + \text{h.c.},
\end{equation}
where $\ell$ and $\ell^{*}$ denote the lepton and excited lepton fields,
and ${W}_{\mu\nu}$ and $B_{\mu\nu}$ are the SU(2) and U(1) gauge fields.
The quantities $g = e / \sin \theta_{W}$ and $g' = e / \cos \theta_{W}$
represent the corresponding electroweak gauge couplings with the Weinberg mixing angle $\theta_{W}$,
and $\tau$ and $Y$ are the generators of the SU(2) and U(1) groups.
The symbols $f$ and $f^{\prime}$ describe the couplings between SM leptons and excited leptons via gauge interactions
and are chosen to be equal to 1 in the model.

Searches for excited leptons have been previously performed by the CMS Collaboration ~\cite{cms-limit-7TeV, cms-limit-8TeV},
but no evidence for their existence was found, excluding $\mlstar<2.5\TeV$ for the case \meqL.
Searches at the LEP~\cite{Buskulic:1996tw, Abreu:1998jw, Abbiendi:1999sa, Achard:2003hd},
HERA~\cite{H1estar}, and Tevatron~\cite{CDFestar,cdfmu,d0,D0estar} colliders, and by the ATLAS Collaboration
at $\sqrt{s}=7\TeV$~\cite{atlas-limit} and 8\TeV~\cite{ATLAS8TeV, atlaslstar}
also found no evidence for the existence of excited leptons,
setting lower mass limits of 2.2 and 2.8\TeV for excited electrons and muons, respectively, for the case \meqL.

\section{The CMS detector}
\label{sec:detector}

The central feature of the CMS apparatus is a superconducting solenoid of 6\unit{m} internal diameter, providing a magnetic field of 3.8\unit{T}. Within the solenoid volume are a silicon pixel and strip tracker, a lead tungstate crystal electromagnetic calorimeter (ECAL), and a brass and scintillator hadron calorimeter (HCAL), each composed of a barrel and two endcap sections. Forward calorimeters extend the pseudorapidity ($\eta$) coverage provided by the barrel and endcap detectors. Muons are detected in gas-ionization chambers embedded in the steel flux-return yoke outside the solenoid.

Events of interest are selected using a two-tiered trigger system~\cite{Khachatryan:2016bia}.
The first level, composed of custom hardware processors, uses information from the calorimeters and muon detectors to select events at a rate of less than 100\unit{kHz} within a time interval of less than 4\mus.
The second level, known as the high-level trigger, consists of a farm of processors running a version of the full event reconstruction software optimized for fast processing, and reduces the event rate to around 1\unit{kHz} before data storage.

A more detailed description of the CMS detector, together with a definition of the coordinate system used and the relevant kinematic variables, can be found in Ref.~\cite{Chatrchyan:2008zzk}.

\section{Data and simulated samples}
\label{sec:Samples}

The data used for this analysis correspond to an integrated luminosity of 35.9\fbinv,
recorded by the CMS detector in $\Pp\Pp$ collisions at a center-of-mass energy of 13\TeV in 2016.
Highly efficient triggers with 98--99\% efficiency are used for this analysis, for events that satisfy the offline selection criteria.
Events for the \eeg channel are selected using double-electron triggers that impose a transverse energy threshold of 33\GeV and online identification criteria for both electron candidates.
For the \mmg channel, events are selected using single-muon triggers with muon isolation criteria and a threshold of 24\GeV
on the transverse momentum \pt of the muon.

To model the detector acceptance and event selection efficiency of signal events,
simulated signal samples are generated with \PYTHIA 8.205~\cite{SJOSTRAND2015159} for excited lepton masses ranging from 0.25 to 5\TeV at intervals of 0.25\TeV up to 4\TeV
and at intervals of 0.5\TeV between 4 and 5\TeV.
The CUETP8M1~\cite{CUETP8M1} underlying-event tune is used for all simulated samples.
The signal samples are generated at the compositeness scale $\Lambda=10\TeV$,
but are also used for different compositeness scale interpretations since
this parameter has a negligible impact on the kinematic distributions of final-state particles.
The simulated signals are generated at leading
order (LO) in perturbative quantum chromodynamics (QCD), and a mass-dependent $K$ factor for
next-to-leading-order (NLO) corrections is applied~\cite{kfactor}.

Major SM background processes such as \Zgamma and \ttgamma are generated at NLO using the \MGvATNLO 2.3.3 generator~\cite{aMCNLO},
while \WW, \WZ, and \ZZ backgrounds are generated at LO with \PYTHIA.
The cross section for \WW production is calculated at next-to-next-to-leading order~\cite{WWNNLO} and the cross sections for \WZ and \ZZ production are computed at NLO~\cite{MCFM66}.

The NNPDF3.0~\cite{Ball:2014uwa} parton distribution function (PDF) set is used for all generated samples.
The generated events are processed through the full \GEANTfour~\cite{Geant} simulation of the CMS detector.
The effect of multiple $\Pp\Pp$ interactions within the same bunch crossing or adjacent bunch crossings (pileup) is
emulated by superimposing minimum bias events on the simulated events.
The simulated events are weighted to match the pileup distribution observed in the data.

\section{Event reconstruction and selection}
\label{sec:objects}

The event selection uses a particle-flow
algorithm~\cite{CMS-PRF-14-001} for optimal efficiency, kinematic resolution and purity in physics object reconstruction and identification.
The particle-flow algorithm aims to reconstruct and identify individual particles in an event
using an optimized combination of information from the various elements of the CMS detector.
The energy of photons is directly obtained from the ECAL measurement.
The energy of electrons is determined from a combination of the electron momentum at the primary interaction vertex as determined by the tracker, the energy of the corresponding ECAL cluster, and the energy sum of all bremsstrahlung photons spatially compatible with originating from the electron track.
The momentum of muons is obtained from the curvature of the corresponding track. The energy of charged hadrons is determined from a combination of their momentum measured in the tracker and the matching ECAL and HCAL energy deposits, corrected for zero-suppression effects and for the response function of the calorimeters to hadronic showers. Finally, the energy of neutral hadrons is obtained from the corresponding corrected ECAL and HCAL energies.

Events must have at least one primary vertex with at least four associated tracks,
with the transverse (longitudinal) position within 2 (24)\unit{cm} from the nominal collision point.
The reconstructed vertex with the largest value of summed physics-object $\pt^2$ is taken to be the primary $\Pp\Pp$ interaction vertex.
Here the physics objects are the jets, clustered using a jet finding algorithm~\cite{Cacciari:2008gp,Cacciari:2011ma} with the tracks assigned to the vertex as inputs, and the associated missing transverse momentum (\ptmiss), taken as the negative vector sum of the \pt of those jets.

Electron candidates are reconstructed by matching clusters of ECAL energy deposits with tracks in the inner tracker.
The momentum resolution ranges from 1.7 to 4.5\% for electrons from $\PZ$ boson decays,
depending on the electron $\eta$ and the bremsstrahlung energy emitted~\cite{CMS:EGM-13-001}.
The electron candidates are required to have $\pt>35\GeV$ and to be within the region $\abs{\eta}<2.5$.
The barrel-endcap transition region $1.44<\abs{\eta}<1.56$ is excluded.
The electron candidates have to pass a set of identification requirements
on the spatial distribution of energy deposits in the ECAL,
on the ratio of the associated HCAL and ECAL energy deposits,
on the isolation in the calorimeters,
on the quality of the matching between the ECAL clusters and the associated track in the inner tracker,
and on the agreement between the energy reconstructed in the ECAL and the momentum of the associated track.

Muon candidates are reconstructed as tracks in the muon detector that are matched to the tracks found in the inner tracker.
The momentum resolution for muons with \pt up to 100\GeV is 1\% in the barrel and 3\% in the endcaps.
The resolution in the barrel is better than 10\% for muons with \pt up to 1\TeV~\cite{muo-16-001}.
The muon candidates must have $\pt>35\GeV$ and be within $\abs{\eta}<2.4$.
The muon candidates are required to pass identification criteria optimized for muons having large \pt~\cite{muo-16-001} and to be isolated in the tracking systems.

Photon reconstruction starts from energy deposits in the ECAL,
and photon candidates that have associated tracks are rejected.
The photon energy resolution ranges from 1 to 5\%, depending on photon $\eta$ and showers in the ECAL~\cite{egm-dps-2014}.
The photon candidates are required to have $\pt>35\GeV$ and $\abs{\eta}<2.5$,
with those in the transition region $1.44<\abs{\eta}<1.56$ being excluded from the analysis.
A multivariate analysis (MVA) technique is used for photon identification, with shower shape variables and photon isolation sums in the ECAL and HCAL as inputs~\cite{egm-dps-2017, egm-dps-2014}.

Events are required to have two SF leptons and a photon.
The two SF leptons are not required to be of opposite-sign charge because
this would result in a signal efficiency loss of a few \% at high lepton \pt, especially for the \eeg channel~\cite{Chatrchyan:2011wq},
whereas the background arising from events with same-sign dileptons is minimal.
The selected leptons must be separated from the photon by $\DR = \sqrt{\smash[b]{(\Delta \eta)^2 + (\Delta \phi)^2}} > 0.7$,
where $\phi$ is the azimuthal angle measured in radians.
In addition, the invariant mass of the two SF leptons \mll is required to be larger than 116\GeV
in order to suppress the dominant background contribution from $\PZ$ boson production (\PZ boson veto criteria).

\section{Background modeling}
\label{sec:background}

The major backgrounds in this search originate from SM processes
with final states consisting of two prompt SF leptons and a prompt photon or a jet misidentified as a photon.
The expected fraction of background events that have a jet misidentified as a lepton is less than 1\% for both channels~\cite{Zprime},
therefore this background source is not considered in the analysis.
Backgrounds associated with a prompt photon are estimated using simulation,
while the yield of those associated with a jet is derived from the data.

The dominant background arises from Drell--Yan process accompanied by a prompt photon (\Zgamma),
which has a signature similar to the signal.
Within the detector acceptance, \PZ boson production dominates the process,
and its contribution is efficiently suppressed by applying the \PZ boson veto criteria described in Section~\ref{sec:objects}.
The fraction of \Zgamma events after the event selection is approximately 70\% of the background.
Another prompt photon background comes from top quark pair production in association with a photon (\ttgamma).
These events form approximately 10\% of the background.
In addition, triboson processes such as \WWgamma, \WZgamma, and \ZZgamma (VV$\Pgg$) also give rise to 5\% of the background.

The other major background contribution consists of events with two prompt leptons and a photon that originates from a jet,
hereafter referred to as the jet background.
The estimation of this background is derived from data in a control region composed of events
with two leptons and a photon passing all the kinematic requirements and lepton identification criteria defined in the event selection but failing the photon identification.
The events in the control region, scaled by a weight factor derived from the misidentification rate of photon candidates from jets,
provide the jet background prediction in the signal region.

The misidentification rate is measured using data in a sideband ($50<\mll<116\GeV$) of the dilepton mass distribution in the signal region ($\mll>116\GeV$).
The photon identification is not required for events in the sideband, thereby the data in the sideband
are enriched with \PZ boson events associated with a photon originating from a jet.
To remove prompt photon contamination in the data,
the distribution of the MVA variable used for the photon identification is employed,
fitting the MVA variable distribution of the data events with template distributions for prompt photons and photons originating from jets.
The MVA distribution of simulated prompt photon background events in the sideband is used as the template distribution for prompt photons.
The template for photons originating from jets is obtained from
a data sample enriched with \PW+jets events, where the $\PW$ boson decays leptonically and a jet supplies a photon candidate.
Events in the sample are required to have a muon of $\pt>35\GeV$, $\ptmiss>35\GeV$,
transverse mass $\sqrt{\smash[b]{2\ptmiss\PT^{\Pgm}[1-\cos\Delta\phi(\ptvecmiss,\ptvec^{\Pgm})]}}$ between 50 and 110\GeV, and a photon candidate.
The misidentification rate is evaluated in several photon \pt bins separately for the barrel and endcaps of ECAL.
The measured misidentification rate ranges from 2 (3)\% at $\pt=35\GeV$ to 0.2 (0.4)\% at $\pt=1\TeV$ in the barrel (endcaps).
In the signal region, the estimated fraction of events arising from background with a jet misidentified as a photon is in the range 5--15\%, depending on the photon \pt.

The kinematic distributions of the parent jets of the misidentified photons in the control region and
those of the signal region are not identical in the same photon \pt range, owing to isolation requirements in the MVA identification.
Consequently, the \pt distributions of leptons, which are correlated with the recoil of the jet in the same event, are different
in the signal region and the control region.
Therefore, an appropriate correction must be applied to the estimate of the lepton \pt distribution
obtained by scaling the control region by a weight factor based on the misidentification rate.
Using the \pt of the dilepton pair as a proxy of the jet recoil,
the dilepton \pt distribution of the jet background estimate
is reweighted to match the correct shape of the dilepton \pt distribution of the jet background events in the signal region.
The shape is directly taken from data in the signal region, removing prompt photon contamination using simulated events.
The reweighting is done for separate photon \pt bins used for the misidentification rate measurement,
in order to retain the yield and the photon \pt spectrum obtained from the previous step.
To validate the estimation procedures, closure tests are performed by applying the same method to derive background on simulated events,
and the observed discrepancies are taken as estimates of the associated systematic uncertainties.
After the correction, the lepton kinematic distributions from the total background prediction
including the corrected jet background obtained from control samples in data are in good agreement with the distributions in data.

\section{Signal modeling}
\label{sec:analysis}

\begin{figure}[hp]
\centering
\includegraphics[width=0.45\textwidth]{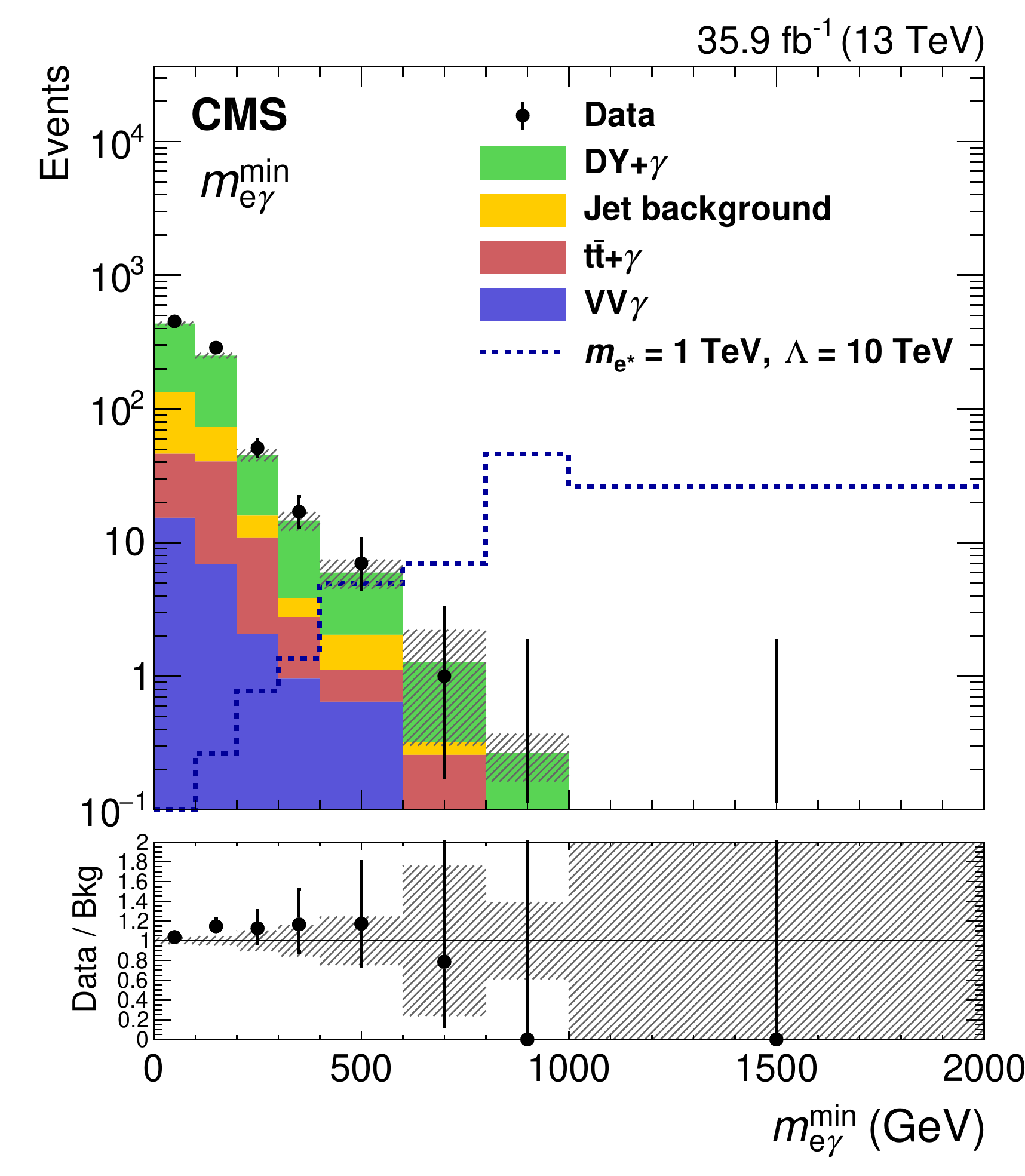}
\includegraphics[width=0.45\textwidth]{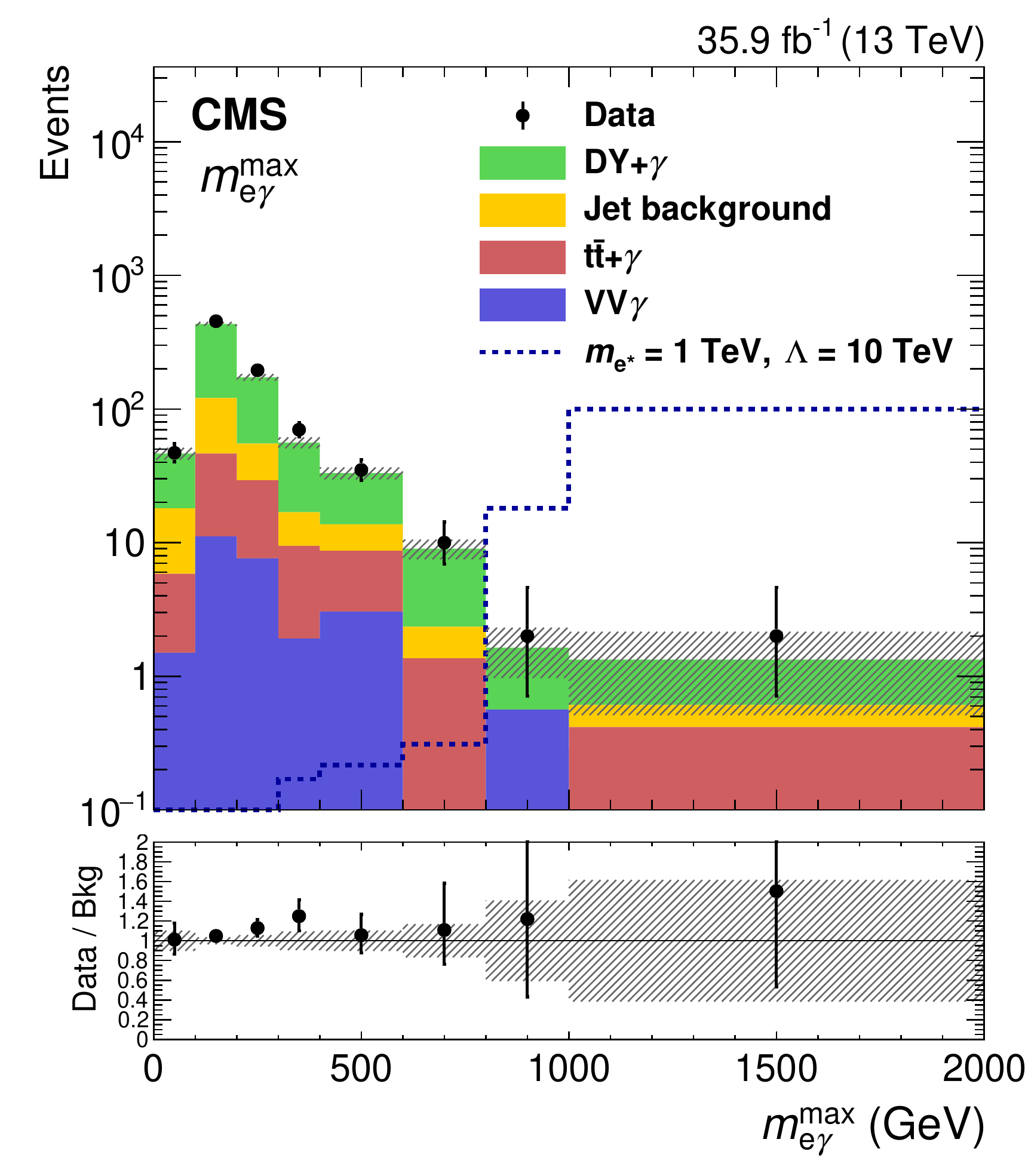}
\includegraphics[width=0.45\textwidth]{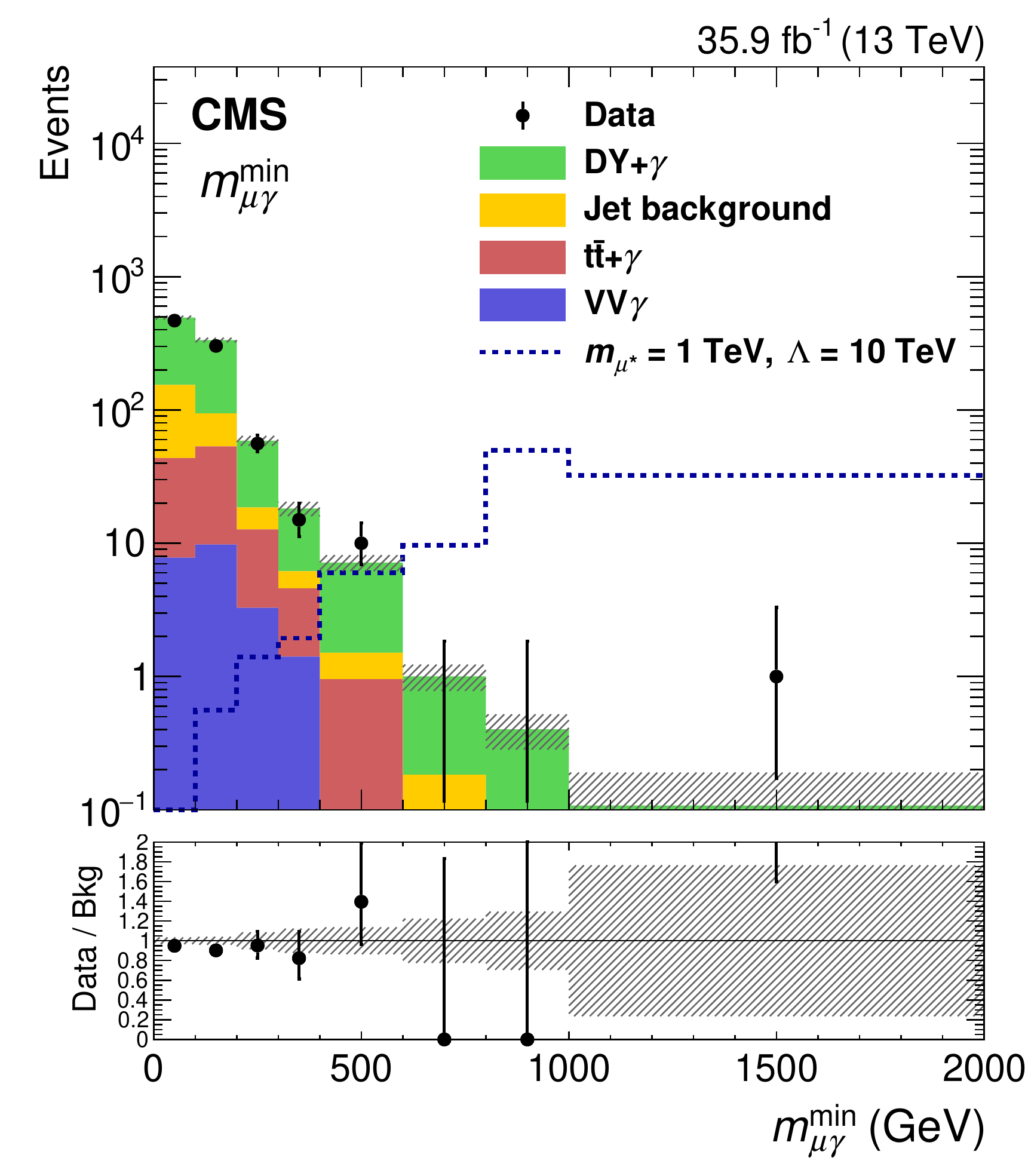}
\includegraphics[width=0.45\textwidth]{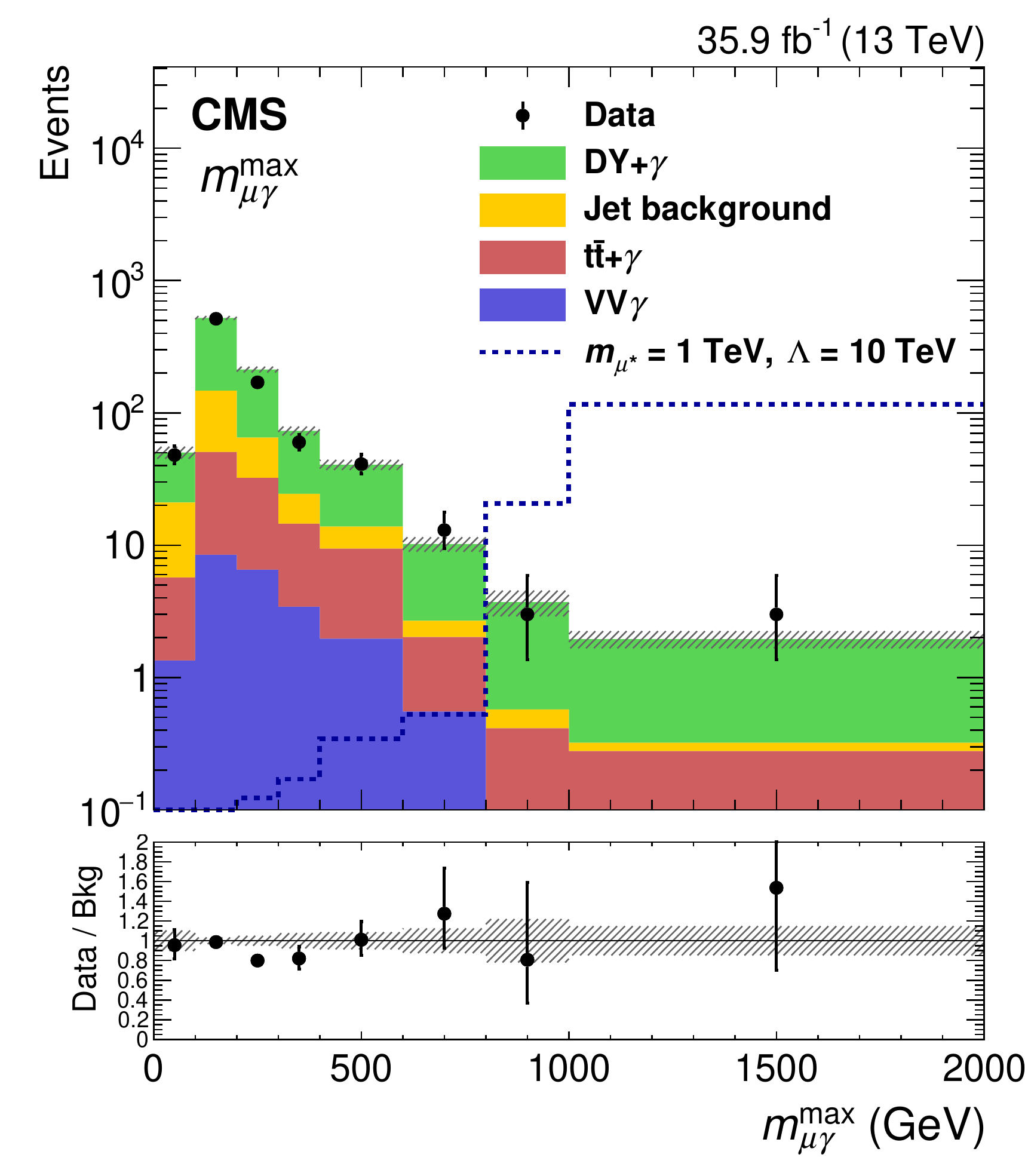}
\caption{
The distributions of \Mmin (left column) and \Mmax (right column)
in the \eeg channel (upper row) and the \mmg channel (lower row).
The points with error bars denote the data
and the stacked histograms show the predictions for each of the backgrounds.
The uncertainty bands of the SM prediction includes only statistical uncertainties.
Signal events for $\mlstar=1\TeV$ at $\Lambda=10\TeV$ are also shown as dotted lines.
The last bin of each distribution includes overflow events.
}
\label{fig:Mmin_Mmax}
\end{figure}

The production of an excited lepton involves two SM leptons in the final state, one from the excited lepton decay and another from CI,
and therefore there are two possible pairings of a lepton with the photon.
The corresponding two invariant masses are referred to as \Mmin and \Mmax.
The \Mmin and \Mmax distributions of observed events along with the background prediction in the signal region are shown in Fig.~\ref{fig:Mmin_Mmax}.

\begin{figure}[h]
\centering
\includegraphics[width=0.45\textwidth]{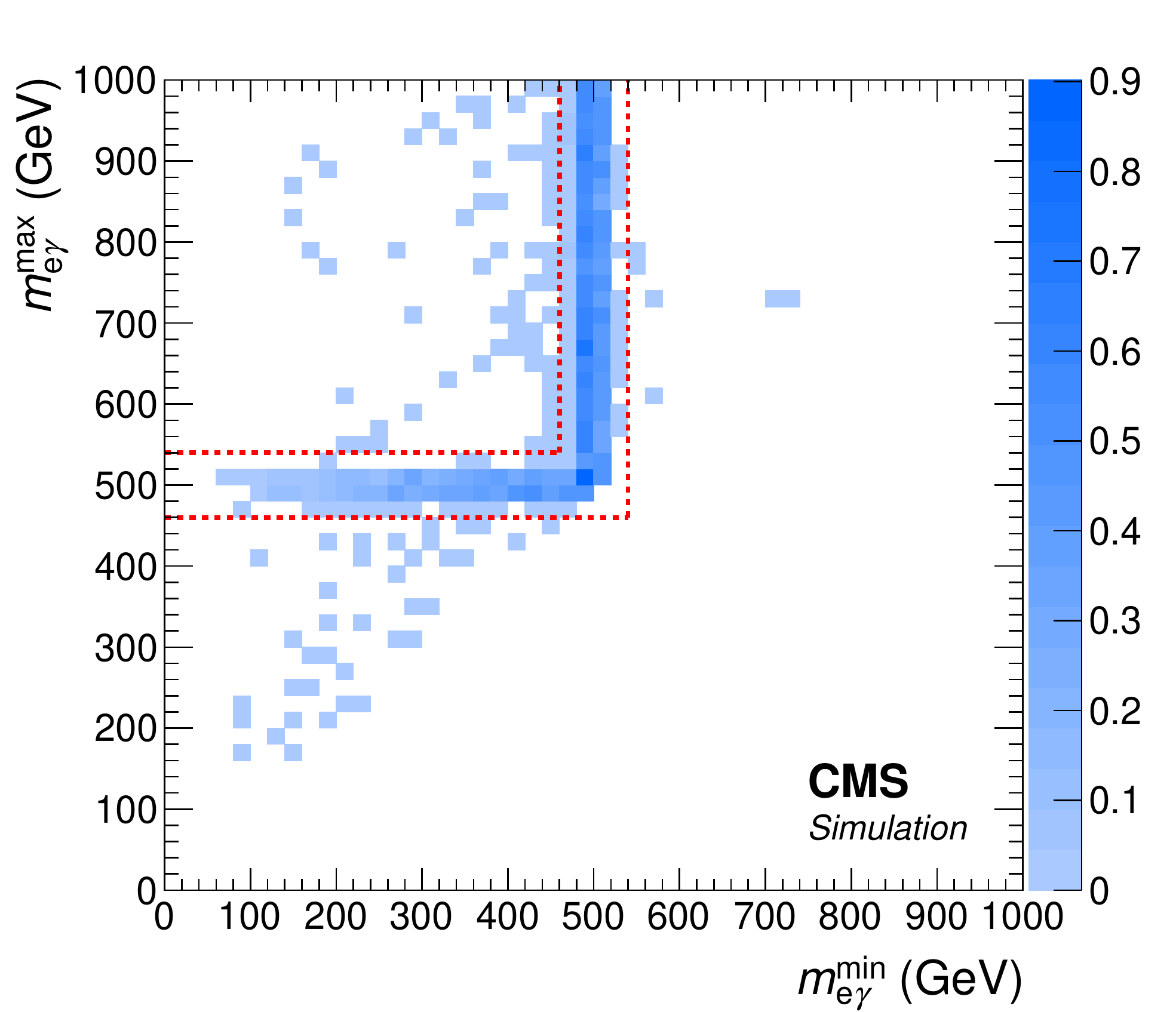}
\includegraphics[width=0.45\textwidth]{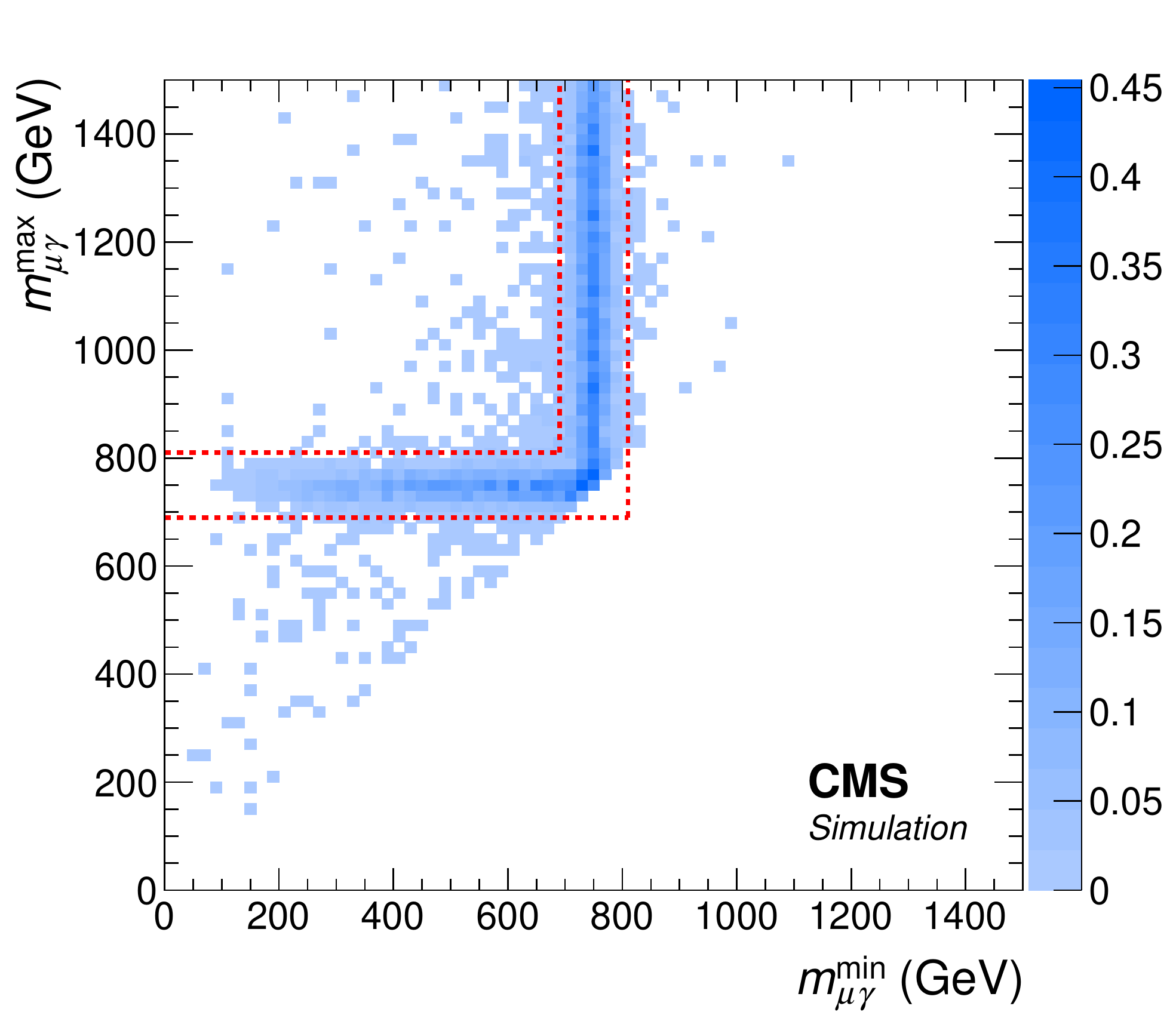}
\caption{The two-dimensional distributions of \Mmax versus \Mmin of excited electrons with a mass of 500\GeV (left) and of excited muons with a mass of 750\GeV (right),
after the event selection, normalized to the expected signal cross section at $\Lambda=10\TeV$. The red dashed lines denote the boundary of the L-shaped search window.
}
\label{fig:2D_Mmin-Mmax_sig}
\end{figure}

\begin{figure}[h]
\centering
\includegraphics[width=0.45\textwidth]{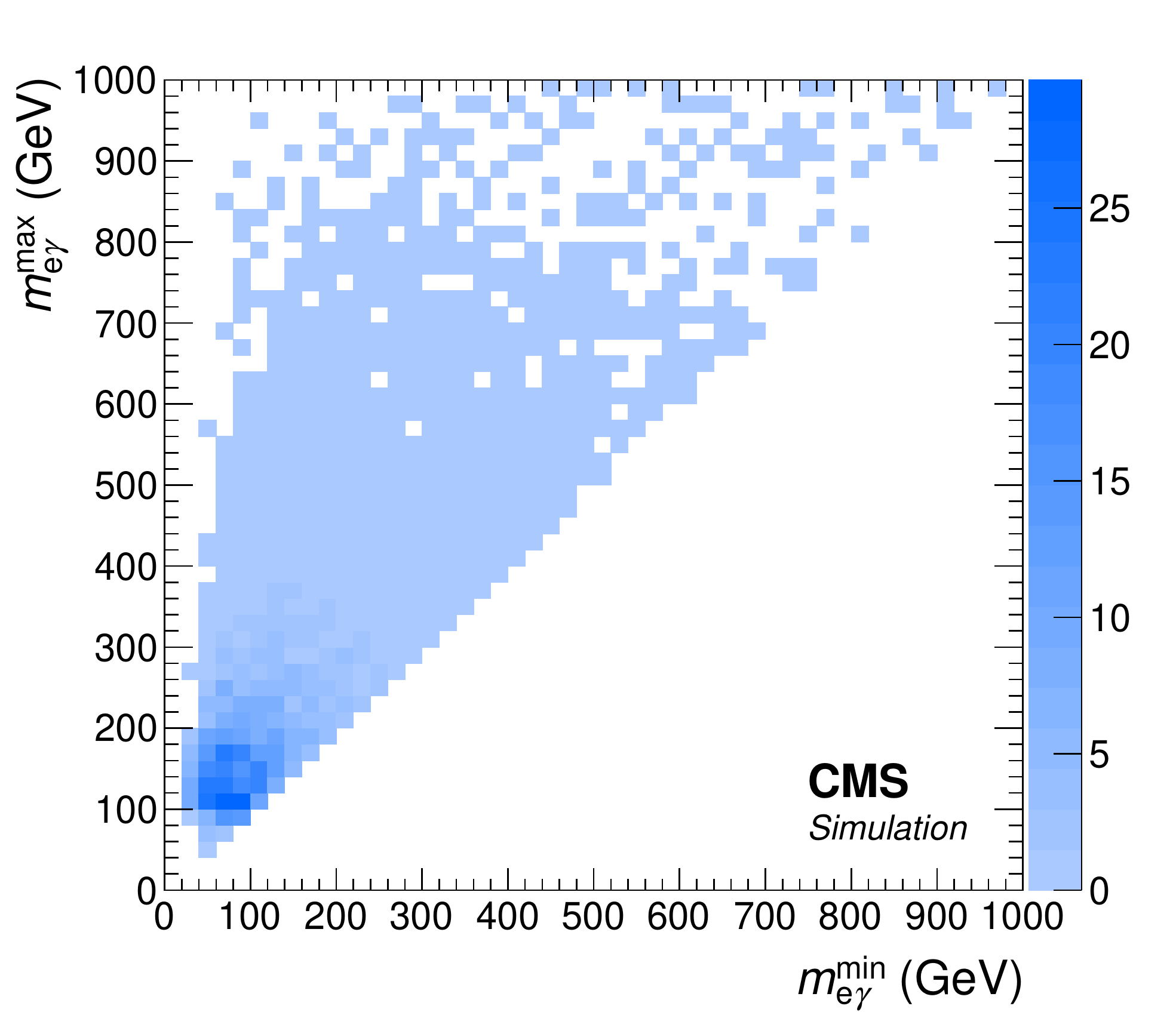}
\includegraphics[width=0.45\textwidth]{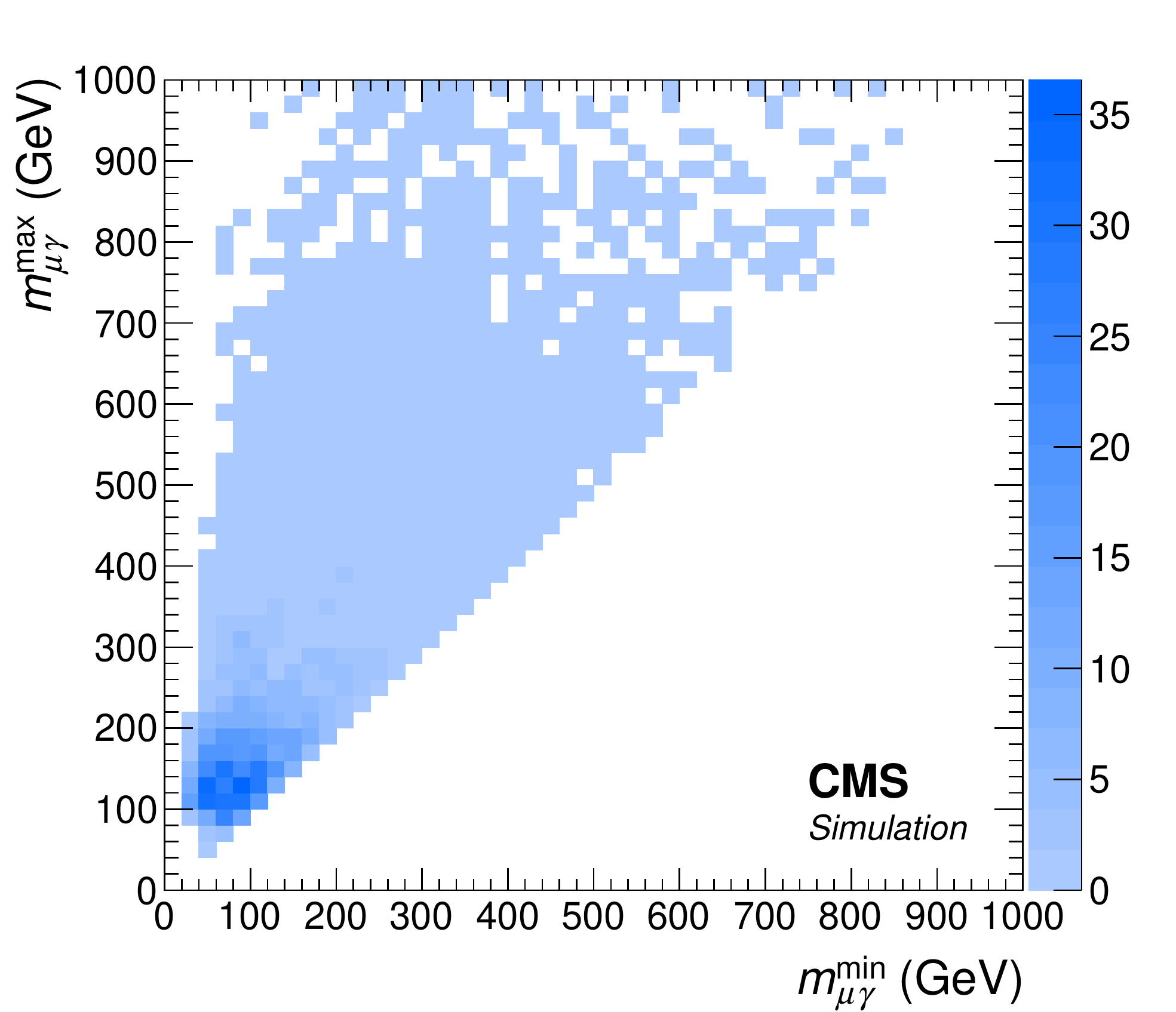}
\caption{The two-dimensional distributions of \Mmax versus \Mmin of \Zgamma background events in the \eeg (left) and \mmg (right) channels, after the event selection, normalized to the cross section for \Zgamma production.
}
\label{fig:2D_Mmin-Mmax_bkg}
\end{figure}

A search window is set in the two-dimensional distribution of \Mmax versus \Mmin.
For \lstar events, either \Mmin or \Mmax corresponds to the reconstructed invariant mass of \lstar.
Therefore, the mass resonance of the signal is concentrated in the shape of a reflected
``L'' as shown in Fig.~\ref{fig:2D_Mmin-Mmax_sig}.
On the other hand, background events have no such correlation in \Mmin and \Mmax
in the low mass region below 1\TeV.
The distribution of the dominant \Zgamma background is shown in Fig.~\ref{fig:2D_Mmin-Mmax_bkg}.
This clear distinction between signal and background events in the distribution of \Mmax versus \Mmin
is used to define L-shaped search windows enhancing the discrimination between signal and background.

For low signal masses, $\mlstar\leq1\TeV$, setting an L-shaped search window significantly improves a discrimination against the background with only a small loss in the signal acceptance.
Therefore, the search window for $\mlstar\leq1\TeV$ is set to be a narrow L-shape centered at \mlstar,
consisting of lower and upper thresholds of \Mmin and \Mmax as shown in Fig.~\ref{fig:2D_Mmin-Mmax_sig}.
The thresholds are optimized for the best expected exclusion limit,
which also provides the optimal discovery potential.
The impact of the energy scale and resolution uncertainties on the signal acceptance is taken into account for the optimization.
The optimized thresholds have values that are within $\pm$8\% of the simulated mass for \estar,
and within ranges that vary from $\pm$4.5\% to $\pm$8\% of the simulated mass for \mustar.

The search windows defined for $\mlstar>1\TeV$, where the background contribution is expected to be negligible,
only impose the lower \Mmax threshold of 1\TeV, thereby maximizing the signal acceptance.

\begin{figure}[!hbtp]
\centering
\includegraphics[width=0.45\textwidth]{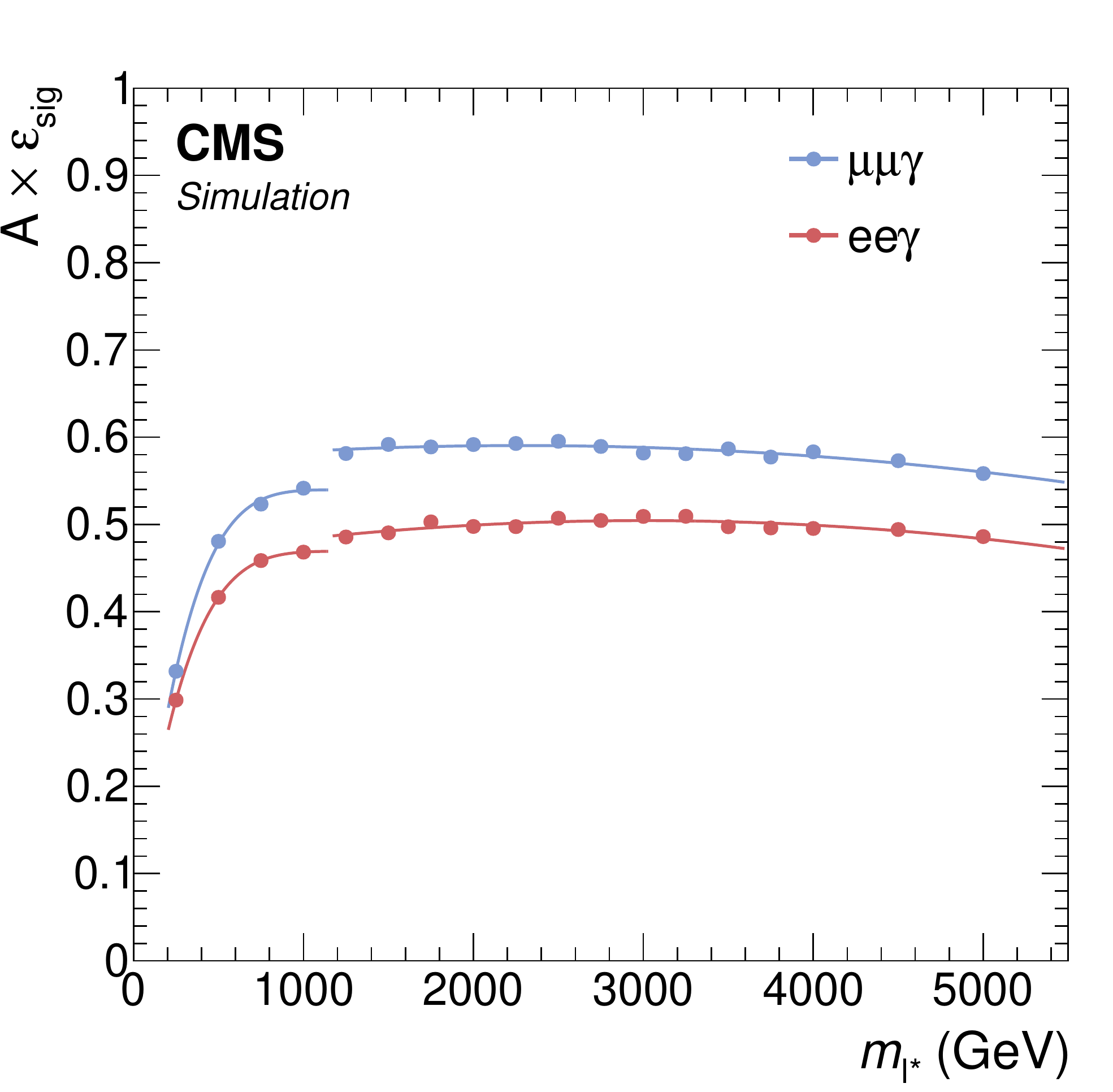}
\caption{
The product of signal acceptance and efficiency as a function of the generated resonance mass
for the \eeg (lower) and \mmg (upper) channels.
Each marker denotes the value measured from the simulated signal sample at a given mass point,
and the lines represent polynomial fits to the measured values.
}
\label{fig:signal_acc_x_eff}
\end{figure}

The product of signal acceptance and efficiency (\Ae) has been measured from the simulated signal samples,
and ranges from 30 to 49\% and from 33 to 59\% for excited electrons and muons, respectively.
The relatively low values of \Ae at low masses mainly result from lepton and photon \pt thresholds.
To determine \Ae for mass points
other than those of the simulated samples, a polynomial fit to the dependence on \mlstar is used for interpolation, as shown in Fig.~\ref{fig:signal_acc_x_eff}.

\section{Systematic uncertainties}
\label{sec:Systematics}

The systematic uncertainties in the signal and the background yields are summarized in Table~\ref{tab:uncsummary}.
The statistical uncertainties in the data and the simulated samples used for the background estimation are dominant.
These uncertainties are expected to be reduced in future by producing simulated samples enriched with high \pt photons.
The statistical uncertainties in the simulated signal samples used to measure \Ae are negligible and therefore not considered in the analysis.
The values of the systematic uncertainties for the jet background estimate coming from control samples in data are also large compared to other systematic uncertainties,
but their impact on the sensitivity of the search is small since the jet background makes up only 5--15\% of the total background.

\begin{table}[!hbtp]
\topcaption{Summary of the systematic uncertainties (in \%) in the signal yield, the prompt photon background prediction, and the jet background prediction.}
\centering
\cmsTable{
\begin{tabular}{l*{6}c}
\hline
\multirow{2}{*}{Source}                    & \multicolumn{3}{c}{\eeg channel (\%)}           & \multicolumn{3}{c}{\mmg channel (\%)}   \\
                                           & Signal  & Prompt $\Pgg$ bkg & Jet bkg         & Signal  & Prompt $\Pgg$ bkg & Jet bkg \\
\hline
Integrated luminosity                      & 2.5     & 2.5                 & \NA             & 2.5     & 2.5                 & \NA     \\
Pileup                                     & 1       & 1                   & \NA             & 1       & 1                   & \NA     \\
Trigger                                    & 1       & 1                   & \NA             & 2       & 2                   & \NA     \\
Lepton efficiency                          & 2.5     & 2.5                 & \NA             & 2       & 2                   & \NA     \\
Photon efficiency                          & 1.5     & 1.5                 & \NA             & 1.5     & 1.5                 & \NA     \\
$\Pe$/$\Pgg$ energy scale \& resolution  & 2       & 2                   & \NA             & 2.5     & 2.5                 & \NA     \\
$\Pgm$ momentum scale \& resolution        & \NA     & \NA                 & \NA             & 2       & 2                   & \NA     \\
PDF \& scales                              & 2       & 10                  & \NA             & 2       & 10                  & \NA     \\
Jet bkg estimate                           & \NA     & \NA                 & 54--90          & \NA     & \NA                 & 54--90  \\
Sample size                                & \NA     & 8--44               & 5--47           & \NA     & 10--25              & 6--58   \\
\hline
\end{tabular}
}
\label{tab:uncsummary}
\end{table}

The integrated luminosity has been measured with a precision of 2.5\%~\cite{LUM-17-001}.
The effect of pileup modeling on the selection efficiency is measured to be less than a percent over all mass points,
and a 1\% uncertainty is assigned for it.
Systematic uncertainties in the signal selection efficiency include uncertainties arising from
the trigger selection, lepton identification~\cite{muo-16-001} and photon identification~\cite{egm-dps-2017, egm-dps-2014}.
Various sources of potential biases in the selection efficiency measurements are considered to determine such uncertainties.
Uncertainties in the \Pe/$\Pgg$ energy scale and resolution,
and muon momentum scale and resolution translate into uncertainties in the signal acceptance.
The effect on the signal acceptance is evaluated by shifting and smearing the \pt of each object by $\pm$1 standard deviation.
Although the fraction of energy affected by the \Pe/$\Pgg$ energy scale and resolution uncertainties is larger for \eeg events than for \mmg events,
because the mass window is narrower and the mass resolution is worse, the impact on the acceptance is measured to be stronger in the latter case.

Systematic uncertainties in the signal acceptance and background cross sections due to the PDF choice
have been estimated by following the PDF4LHC prescription~\cite{PDF4LHC}.
Renormalization and factorization scale uncertainties are evaluated by varying the scales up and down by a factor of 2, both simultaneously and independently.
The maximum change observed among the corresponding variations in the signal and the background yields is taken as an estimate of the associated systematic uncertainty.
Uncertainties of 2 and 10\% are assigned for the PDF and scale uncertainties in the signal acceptance and the background yields, respectively.

For the jet background estimate, the following uncertainties are taken into account:
statistical uncertainties in the jet-to-photon misidentification rate measurement (8--72\%),
systematic uncertainties in the template distributions used for the misidentification rate measurement (20\%),
and systematic uncertainties based on the discrepancies observed in the closure tests
for the jet background determination procedure performed on simulated samples (50\%).
The total systematic uncertainty in the jet background prediction is evaluated by summing these uncertainties in quadrature and amounts to 54\%, increasing to 90\% for masses above 1\TeV.

\section{Results}
\label{sec:results}

Within the uncertainties, the data are found to be consistent with the background prediction as summarized in Tables~\ref{tab:eeg_yields} and \ref{tab:mmg_yields}.
We set 95\% confidence level (\CL) upper limits on the production cross sections of excited electrons and excited muons,
and corresponding lower limits on the compositeness scale $\Lambda$, as a function of the excited lepton mass
using a single-bin counting method \cite{higgstool}.
The limits are computed with the modified frequentist \CLs method~\cite{Junk:1999kv, CLs}, with a likelihood ratio used as a test statistic.
The systematic uncertainties are treated as nuisance parameters with log-normal priors.
The limits are set in the mass range between 0.25 and 5.0\TeV.

\begin{table}[!hbtp]
\centering
\topcaption{The observed yield and the SM prediction in the search window of the given \mlstar in the \eeg channel.
The symbols \Ndata, \Nprompt, and \Njet represent the number of events in data, the prompt photon background prediction, and the jet background estimate, respectively, together with statistical and systematical uncertainties.
}
\label{tab:eeg_yields}
\begin{tabular}{cccccccc}
\hline
\melstar (\GeVns) & Window (\GeVns) & \Ndata & \Nprompt & \Njet & \Ae \\
\hline
250 & 230--270 & 84 & 74.4 $\pm$ 6.8 $\pm$ 8.0 & 12.5 $\pm$ 0.7 $\pm$ 6.9 & 0.30 \\
275 & 253--297 & 80 & 50.9 $\pm$ 6.0 $\pm$ 5.4 & 10.0 $\pm$ 0.6 $\pm$ 5.5 & 0.32 \\
300 & 276--324 & 68 & 44.7 $\pm$ 5.5 $\pm$ 4.8 & 7.6 $\pm$ 0.5 $\pm$ 4.2 & 0.33 \\
330 & 304--356 & 51 & 40.4 $\pm$ 4.7 $\pm$ 4.3 & 5.9 $\pm$ 0.5 $\pm$ 3.3 & 0.35 \\
360 & 331--389 & 39 & 28.1 $\pm$ 3.8 $\pm$ 3.0 & 4.0 $\pm$ 0.4 $\pm$ 2.2 & 0.36 \\
400 & 368--432 & 27 & 19.4 $\pm$ 3.0 $\pm$ 2.1 & 3.3 $\pm$ 0.3 $\pm$ 1.8 & 0.38 \\
450 & 414--486 & 17 & 15.8 $\pm$ 2.5 $\pm$ 1.7 & 2.8 $\pm$ 0.4 $\pm$ 1.6 & 0.40 \\
500 & 460--540 & 16 & 12.3 $\pm$ 1.9 $\pm$ 1.3 & 2.3 $\pm$ 0.3 $\pm$ 1.3 & 0.42 \\
550 & 506--594 & 15 & 8.2 $\pm$ 1.7 $\pm$ 0.9 & 1.6 $\pm$ 0.2 $\pm$ 0.9 & 0.43 \\
600 & 552--648 & 10 & 7.6 $\pm$ 1.8 $\pm$ 0.8 & 1.2 $\pm$ 0.2 $\pm$ 0.7 & 0.44 \\
650 & 598--702 & 6 & 4.9 $\pm$ 1.3 $\pm$ 0.5 & 0.8 $\pm$ 0.2 $\pm$ 0.5 & 0.45 \\
700 & 644--756 & 9 & 3.6 $\pm$ 1.4 $\pm$ 0.4 & 0.5 $\pm$ 0.1 $\pm$ 0.3 & 0.45 \\
750 & 690--810 & 5 & 3.4 $\pm$ 1.3 $\pm$ 0.4 & 0.3 $\pm$ 0.1 $\pm$ 0.2 & 0.46 \\
800 & 736--864 & 1 & 2.9 $\pm$ 1.1 $\pm$ 0.3 & 0.3 $\pm$ 0.1 $\pm$ 0.2 & 0.46 \\
900 & 828--972 & 1 & 1.5 $\pm$ 0.6 $\pm$ 0.2 & 0.1 $\pm$ 0.1 $\pm$ 0.1 & 0.47 \\
1000 & 920--1080 & 1 & 0.8 $\pm$ 0.8 $\pm$ 0.1 & 0.1 $\pm$ 0.1 $\pm$ 0.1 & 0.47 \\
$>1000$ & $\geq1058$ & 1 & 1.4 $\pm$ 0.5 $\pm$ 0.2 & 0.1 $\pm$ 0.1 $\pm$ 0.1 & 0.49 \\
\hline
\end{tabular}
\end{table}

\begin{table}[!hbtp]
\centering
\topcaption{The observed yield and the SM prediction in the search window of the given \mlstar in the \mmg channel.
The symbols \Ndata, \Nprompt, and \Njet represent the number of events in data, the prompt photon background prediction, and the jet background estimate, respectively, together with statistical and systematical uncertainties.
}
\label{tab:mmg_yields}
\begin{tabular}{cccccccc}
\hline
\mmulstar (\GeVns) & Window (\GeVns) & \Ndata & \Nprompt & \Njet & \Ae \\
\hline
250 & 238--262 & 41 & 43.8 $\pm$ 4.9 $\pm$ 4.9 & 8.7 $\pm$ 0.6 $\pm$ 4.8 & 0.33 \\
275 & 261--289 & 38 & 42.8 $\pm$ 5.0 $\pm$ 4.8 & 6.8 $\pm$ 0.5 $\pm$ 3.8 & 0.35 \\
300 & 284--316 & 47 & 35.4 $\pm$ 4.6 $\pm$ 4.0 & 6.5 $\pm$ 0.6 $\pm$ 3.6 & 0.37 \\
330 & 312--348 & 23 & 33.1 $\pm$ 3.9 $\pm$ 3.7 & 5.1 $\pm$ 0.5 $\pm$ 2.8 & 0.39 \\
360 & 340--380 & 24 & 25.8 $\pm$ 3.0 $\pm$ 2.9 & 3.7 $\pm$ 0.4 $\pm$ 2.0 & 0.41 \\
400 & 376--424 & 26 & 22.8 $\pm$ 3.0 $\pm$ 2.6 & 2.2 $\pm$ 0.3 $\pm$ 1.2 & 0.44 \\
450 & 422--478 & 17 & 15.1 $\pm$ 2.3 $\pm$ 1.7 & 1.8 $\pm$ 0.3 $\pm$ 1.0 & 0.46 \\
500 & 467--533 & 14 & 9.8 $\pm$ 1.6 $\pm$ 1.1 & 1.8 $\pm$ 0.3 $\pm$ 1.0 & 0.48 \\
550 & 512--588 & 11 & 10.8 $\pm$ 1.8 $\pm$ 1.2 & 1.0 $\pm$ 0.2 $\pm$ 0.5 & 0.49 \\
600 & 556--644 & 8 & 5.8 $\pm$ 1.2 $\pm$ 0.7 & 0.7 $\pm$ 0.2 $\pm$ 0.4 & 0.51 \\
650 & 600--700 & 10 & 6.8 $\pm$ 1.1 $\pm$ 0.8 & 0.6 $\pm$ 0.1 $\pm$ 0.3 & 0.52 \\
700 & 644--756 & 5 & 5.8 $\pm$ 1.0 $\pm$ 0.7 & 0.4 $\pm$ 0.1 $\pm$ 0.2 & 0.52 \\
750 & 690--810 & 6 & 5.1 $\pm$ 1.0 $\pm$ 0.6 & 0.3 $\pm$ 0.1 $\pm$ 0.2 & 0.53 \\
800 & 736--864 & 3 & 4.5 $\pm$ 1.0 $\pm$ 0.5 & 0.2 $\pm$ 0.1 $\pm$ 0.1 & 0.53 \\
900 & 828--972 & 2 & 3.1 $\pm$ 0.8 $\pm$ 0.4 & 0.1 $\pm$ 0.1 $\pm$ 0.1 & 0.54 \\
1000 & 920--1080 & 0 & 1.1 $\pm$ 0.3 $\pm$ 0.1 & 0.1 $\pm$ 0.0 $\pm$ 0.1 & 0.54 \\
$>1000$ & $\geq1058$ & 3 & 1.5 $\pm$ 0.3 $\pm$ 0.2 & 0.0 $\pm$ 0.0 $\pm$ 0.0 & 0.59 \\
\hline
\end{tabular}
\end{table}

\begin{figure}[!hbtp]
\centering
\includegraphics[width=0.45\textwidth]{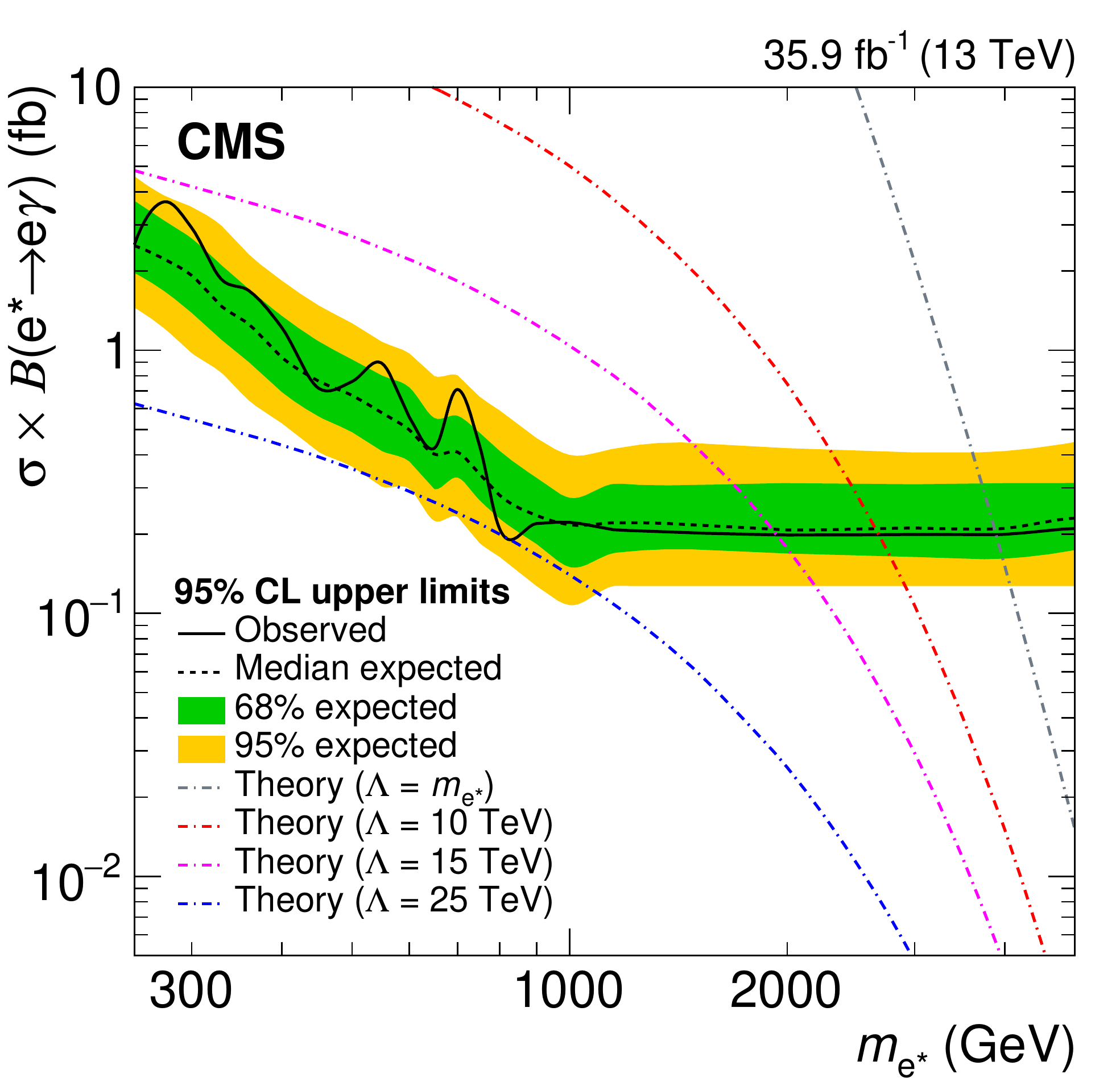}
\includegraphics[width=0.45\textwidth]{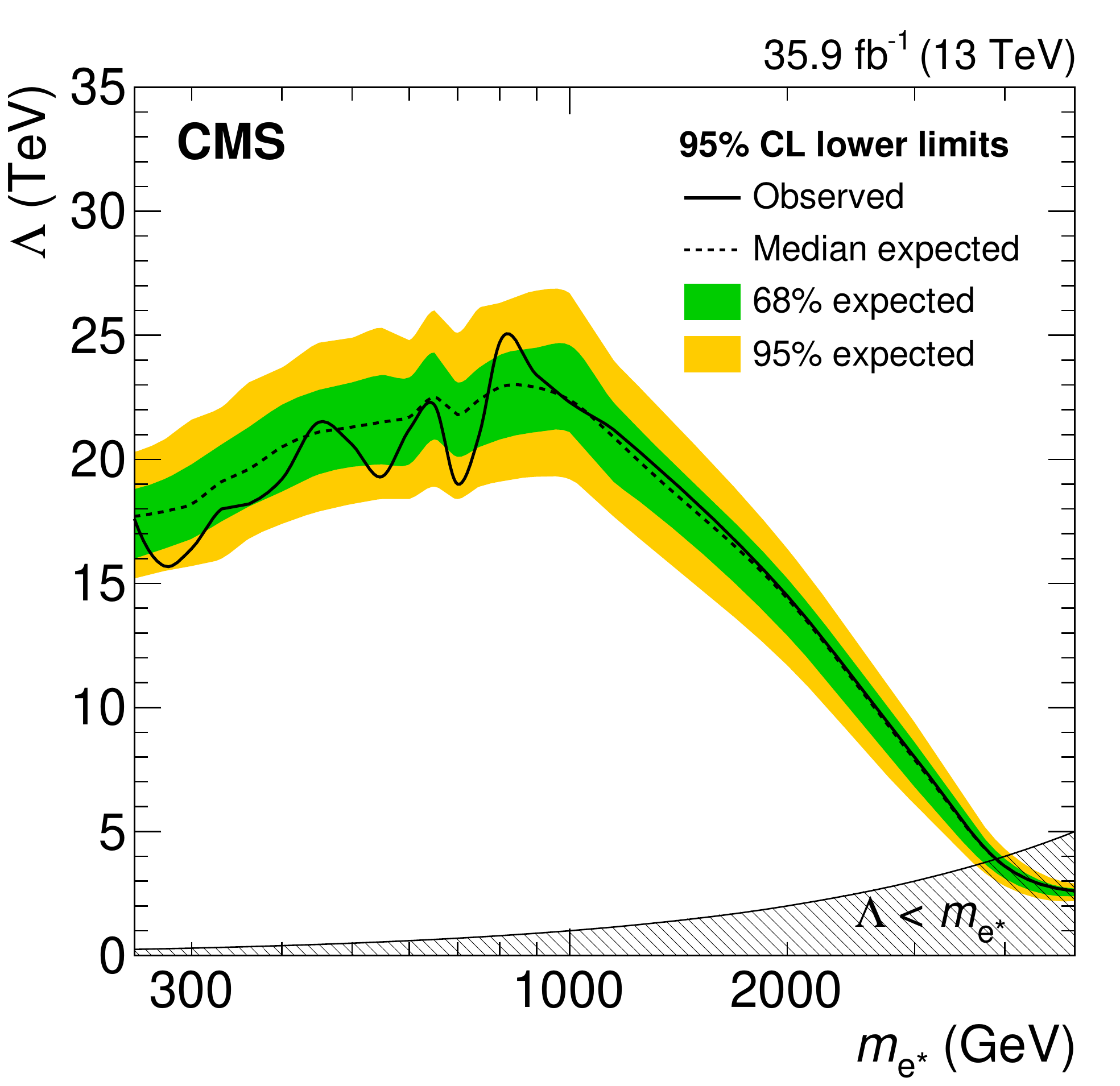}
\includegraphics[width=0.45\textwidth]{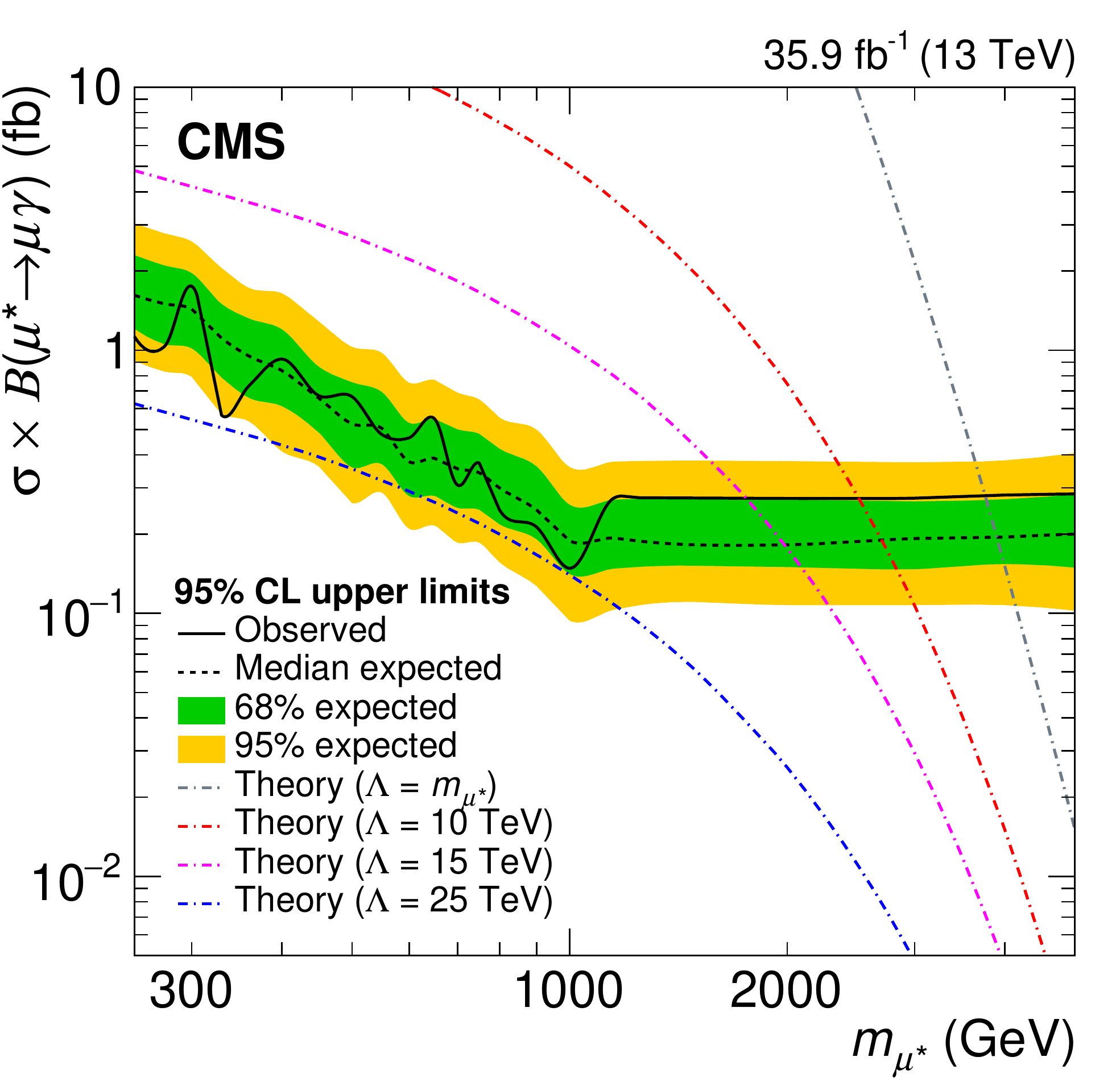}
\includegraphics[width=0.45\textwidth]{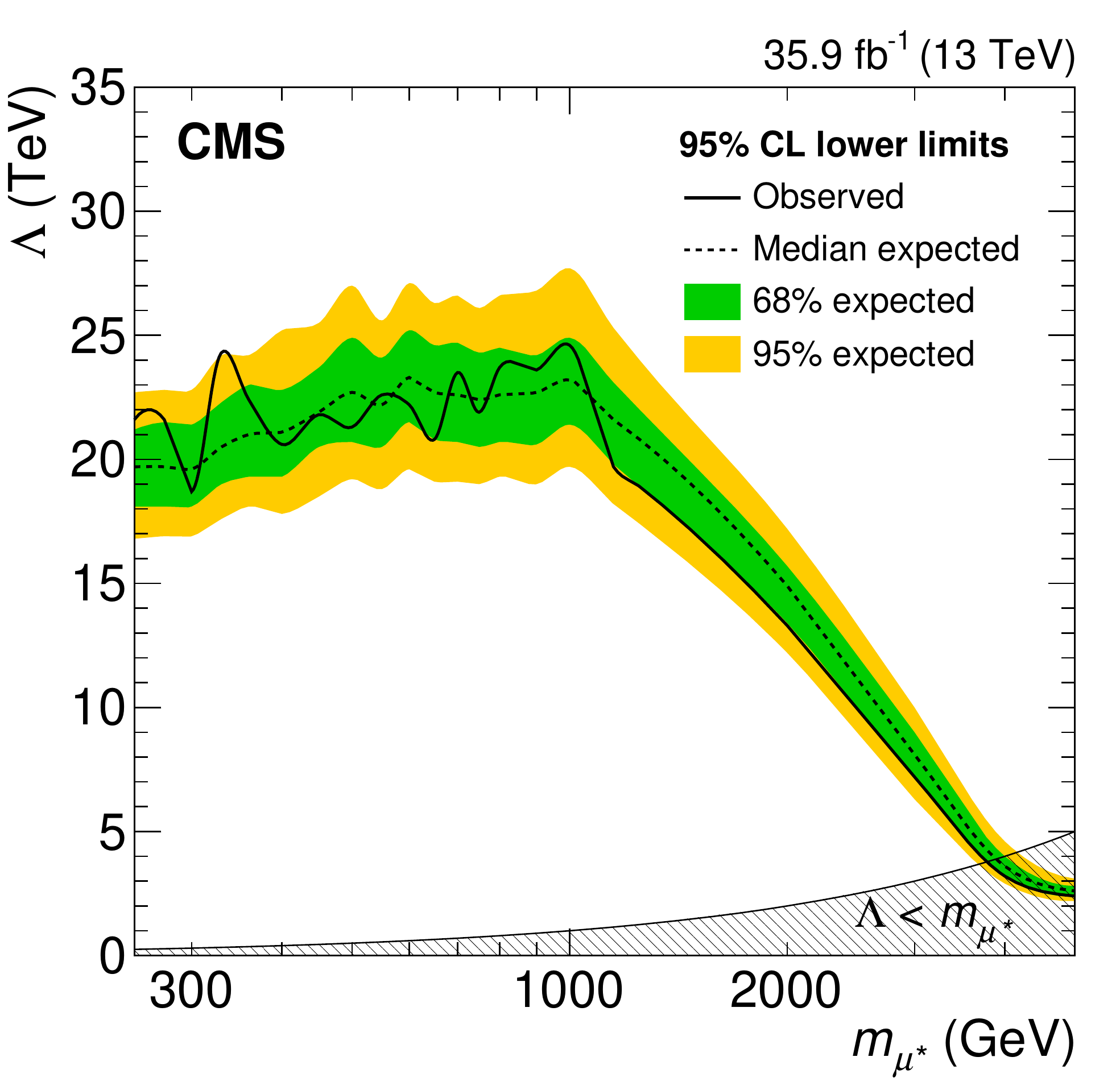}
\caption{
Observed~(solid) and
expected~(dashed)  95\% \CL upper limits
on the product of the production cross section and branching fraction (left column) and lower limits on the compositeness scale (right column)
as a function of signal mass \mlstar,
together with the 68\%~(green, inner) and 95\%~(yellow, outer) quantiles of the expected limit, for \estar (upper row) and \mustar (lower row).
}
\label{fig:limits}
\end{figure}

Figure~\ref{fig:limits} shows the 95\% \CL upper limits
on the product of the signal cross section and branching fraction to \llg final states,
$\sigma\,\mathcal{B}(\lstartollg)$ and lower limits on $\Lambda$, as a function of the resonance mass.
The observed limits are denoted by the solid black lines and the expected limits for the background-only hypothesis
are represented by the dashed black lines. The 68 and 95\% \CL ranges
are shown with the green and yellow bands, respectively.
The observed limits on the signal cross section range from 3.7 to 0.2\unit{fb} as a function of \mlstar and are consistent with the expected limits.
The fluctuations of the expected limits and the uncertainty bands, which appear for $\mlstar<1\TeV$,
are consequences of the statistical fluctuations from the limited number of simulated events passing the event selection.
The dashed lines in the left plots in Fig.~\ref{fig:limits} represent the theoretical cross sections including the NLO QCD correction factors
for \meqL,~10, 15, and 25\TeV, in the sequence on the plot from bottom to top.

The lower limit on \mlstar depends on $\Lambda$ since both \mlstar and $\Lambda$ are free parameters in the reference model.
The observation excludes $\mlstar < 3.9$ (3.8)\TeV for excited electrons (muons) in the case where \meqL.
The exclusion on $\Lambda$ ranges from 15 to 25\TeV in the mass range \mlstar between 0.25 and 1.0\TeV,
and decreases with increasing mass up to approximately 4\TeV for $\mlstar>1.0\TeV$.
The best observed limit on $\Lambda$ is obtained in the mass range between 0.5 and 1.0\TeV, excluding $\Lambda$ below 25\TeV for both excited electrons and muons.
A summary of representative exclusion limits is given in Table~\ref{tab:limits}.

\begin{table}[!hbtp]
\topcaption{Summary of the observed (expected) lower limits on \mlstar, assuming \meqL, and the best observed (expected) lower limits on $\Lambda$ in the mass range 0.5--1.0\TeV.}
\centering
\begin{tabular}{ccc}
\hline
\multirow{2}{*}{Channel} & Observed (expected) limit & Observed (expected) limit \\
                         & on \mlstar for $\mlstar = \Lambda$, \TeV &  on $\Lambda$ for $\mlstar\approx1\TeV$, \TeV            \\
\hline
\eeg                     & 3.9 (3.8)                     & 25 (23)                        \\
\mmg                     & 3.8 (3.9)                     & 25 (23)                        \\
\hline
\end{tabular}
\label{tab:limits}
\end{table}

\section{Summary}

A search has been presented for excited electrons and muons in \llg final states at the LHC.
The search is based on a data sample corresponding to an integrated luminosity of 35.9\fbinv
of proton-proton collisions at a center-of-mass energy of 13\TeV, collected with the CMS detector in 2016.
No significant excess over the standard model prediction is observed in the data,
and 95\% confidence level upper and lower limits are set on the product of the signal production cross section and branching fraction to \llg final states
and the compositeness scale, respectively, as a function of the excited lepton mass.
The observed limits on the product of the signal cross section and the branching fraction range from 3.7 to 0.2\unit{fb} as a function of \mlstar.
Excited electrons and muons are excluded for masses below 3.9 and 3.8\TeV, respectively, under the assumption
that the excited lepton mass equals the compositeness scale.
The best observed limit on the compositeness scale is obtained with an excited lepton mass of around 1.0\TeV,
excluding a compositeness scale below 25\TeV for both excited electrons and muons.
These are the first results of a search at $\sqrt{s} = 13\TeV$ for excited leptons
and also the most stringent limits on the excited lepton mass and the compositeness scale to date.

\clearpage

\begin{acknowledgments}

We congratulate our colleagues in the CERN accelerator departments for the excellent performance of the LHC and thank the technical and administrative staffs at CERN and at other CMS institutes for their contributions to the success of the CMS effort. In addition, we gratefully acknowledge the computing centers and personnel of the Worldwide LHC Computing Grid for delivering so effectively the computing infrastructure essential to our analyses. Finally, we acknowledge the enduring support for the construction and operation of the LHC and the CMS detector provided by the following funding agencies: BMBWF and FWF (Austria); FNRS and FWO (Belgium); CNPq, CAPES, FAPERJ, FAPERGS, and FAPESP (Brazil); MES (Bulgaria); CERN; CAS, MoST, and NSFC (China); COLCIENCIAS (Colombia); MSES and CSF (Croatia); RPF (Cyprus); SENESCYT (Ecuador); MoER, ERC IUT, and ERDF (Estonia); Academy of Finland, MEC, and HIP (Finland); CEA and CNRS/IN2P3 (France); BMBF, DFG, and HGF (Germany); GSRT (Greece); NKFIA (Hungary); DAE and DST (India); IPM (Iran); SFI (Ireland); INFN (Italy); MSIP and NRF (Republic of Korea); MES (Latvia); LAS (Lithuania); MOE and UM (Malaysia); BUAP, CINVESTAV, CONACYT, LNS, SEP, and UASLP-FAI (Mexico); MOS (Montenegro); MBIE (New Zealand); PAEC (Pakistan); MSHE and NSC (Poland); FCT (Portugal); JINR (Dubna); MON, RosAtom, RAS, RFBR, and NRC KI (Russia); MESTD (Serbia); SEIDI, CPAN, PCTI, and FEDER (Spain); MOSTR (Sri Lanka); Swiss Funding Agencies (Switzerland); MST (Taipei); ThEPCenter, IPST, STAR, and NSTDA (Thailand); TUBITAK and TAEK (Turkey); NASU and SFFR (Ukraine); STFC (United Kingdom); DOE and NSF (USA).

\hyphenation{Rachada-pisek} Individuals have received support from the Marie-Curie program and the European Research Council and Horizon 2020 Grant, contract No. 675440 (European Union); the Leventis Foundation; the A. P. Sloan Foundation; the Alexander von Humboldt Foundation; the Belgian Federal Science Policy Office; the Fonds pour la Formation \`a la Recherche dans l'Industrie et dans l'Agriculture (FRIA-Belgium); the Agentschap voor Innovatie door Wetenschap en Technologie (IWT-Belgium); the F.R.S.-FNRS and FWO (Belgium) under the ``Excellence of Science - EOS" - be.h project n. 30820817; the Ministry of Education, Youth and Sports (MEYS) of the Czech Republic; the Lend\"ulet (``Momentum") Program and the J\'anos Bolyai Research Scholarship of the Hungarian Academy of Sciences, the New National Excellence Program \'UNKP, the NKFIA research grants 123842, 123959, 124845, 124850 and 125105 (Hungary); the Council of Science and Industrial Research, India; the HOMING PLUS program of the Foundation for Polish Science, cofinanced from European Union, Regional Development Fund, the Mobility Plus program of the Ministry of Science and Higher Education, the National Science Center (Poland), contracts Harmonia 2014/14/M/ST2/00428, Opus 2014/13/B/ST2/02543, 2014/15/B/ST2/03998, and 2015/19/B/ST2/02861, Sonata-bis 2012/07/E/ST2/01406; the National Priorities Research Program by Qatar National Research Fund; the Programa Estatal de Fomento de la Investigaci{\'o}n Cient{\'i}fica y T{\'e}cnica de Excelencia Mar\'{\i}a de Maeztu, grant MDM-2015-0509 and the Programa Severo Ochoa del Principado de Asturias; the Thalis and Aristeia programs cofinanced by EU-ESF and the Greek NSRF; the Rachadapisek Sompot Fund for Postdoctoral Fellowship, Chulalongkorn University and the Chulalongkorn Academic into Its 2nd Century Project Advancement Project (Thailand); the Welch Foundation, contract C-1845; and the Weston Havens Foundation (USA).

\end{acknowledgments}

\clearpage

\bibliography{auto_generated}
\cleardoublepage \appendix\section{The CMS Collaboration \label{app:collab}}\begin{sloppypar}\hyphenpenalty=5000\widowpenalty=500\clubpenalty=5000\vskip\cmsinstskip
\textbf{Yerevan Physics Institute, Yerevan, Armenia}\\*[0pt]
A.M.~Sirunyan, A.~Tumasyan
\vskip\cmsinstskip
\textbf{Institut f\"{u}r Hochenergiephysik, Wien, Austria}\\*[0pt]
W.~Adam, F.~Ambrogi, E.~Asilar, T.~Bergauer, J.~Brandstetter, M.~Dragicevic, J.~Er\"{o}, A.~Escalante~Del~Valle, M.~Flechl, R.~Fr\"{u}hwirth\cmsAuthorMark{1}, V.M.~Ghete, J.~Hrubec, M.~Jeitler\cmsAuthorMark{1}, N.~Krammer, I.~Kr\"{a}tschmer, D.~Liko, T.~Madlener, I.~Mikulec, N.~Rad, H.~Rohringer, J.~Schieck\cmsAuthorMark{1}, R.~Sch\"{o}fbeck, M.~Spanring, D.~Spitzbart, A.~Taurok, W.~Waltenberger, J.~Wittmann, C.-E.~Wulz\cmsAuthorMark{1}, M.~Zarucki
% ...
\vskip\cmsinstskip
% ...
1:  Also at Vienna University of Technology, Vienna, Austria\\
% ...
\end{sloppypar}
\end{document}